\documentclass[preprint,11pt]{JHEP3}

\JHEPspecialurl{http://jhep.sissa.it/JOURNAL/JHEP3.tar.gz}

\usepackage{epsfig,multicol,amsmath}

\usepackage{bm}  
\usepackage{amsmath}     

\usepackage{graphicx}
\usepackage{rotating}
\usepackage{pifont}
\usepackage{relsize}
\usepackage{mathtools}
\usepackage{cite}

\def\beq{\begin{equation}}
\def\eeq{\end{equation}}
\def\bea{\begin{eqnarray}}
\def\eea{\end{eqnarray}}

\def\eq#1{{Eq.~(\ref{#1})}}
\def\fig#1{{Fig.~\ref{#1}}}
\newcommand{\bas}{\bar{\alpha}_S}
\newcommand{\as}{\alpha_S}

\newcommand{\Lb}{\left(}
\newcommand{\Rb}{\right)}
\newcommand{\h}{\frac{1}{2}}

\newcommand{\prm}{^{\,\prime}}
\newcommand{\dpr}{^{\,\prime\prime}}
\newcommand{\He}{=\hspace{0.3cm}}
\newcommand{\he}{\hspace{0.3cm}=\hspace{0.3cm}}

\parskip=\itemsep               

\newcommand{\nn}{\nonumber}
\newcommand{\D}{\partial}

\newcommand{\ga}{\gamma}
\newcommand{\de}{\delta}

\newcommand{\Ga}{\Gamma}
\newcommand{\om}{\omega}

\newcommand{\f}{\frac}
\newcommand{\lab}{\label}

\newcommand{\ml}{\mathlarger}

\title{BFKL Pomeron calculus: nucleus-nucleus scattering }
\author{\Large
Carlos  ~Contreras${}^{a}$\thanks{Email: carlos.contreras@usm.cl},\,\,\,\,Eugene\, Levin${}^{a, b}$ \thanks{Email: leving@post.tau.ac.il., eugeny.levin@usm.cl}\,\,\,and \,\, Jeremy\,S.\,Miller${}^{b, c}$
\thanks{Email:  jeremy.miller@ist.utl.pt}
\\
${}^a$\, Departamento de F\'\i sica, Universidad T\'ecnica
Federico Santa Mar\'\i a, Avda. Espa\~na 1680 and  Centro Cient\'ifico-Tecnol$\acute{o}$gico de Valpara\'\i so, Casilla 110-V,  Valparaiso, Chile\\
${}^b$ \, Department of Particle Physics, School of Physics and Astronomy,
Tel Aviv University, Tel Aviv, 69978, Israel\\
${}^c$\,CENTRA, Departamento de F\'\i sica, Instituto Superior T\'ecnico (IST), Av. Rovisco Pais, 1049-001 Lisboa, Portugal
}

\abstract{In this paper the action of the BFKL Pomeron calculus is re-written in momentum representation,
 and the equations of motion  for nucleus-nucleus collisions are derived, in this representation. 
We found the semi-classical solutions to these equations, outside of the saturation domain.
Inside this domain these equations reduce to the set of delay differential equations,
and their asymptotic solutions are derived. }

\dedicated{PACS: 13.85.-t, 13.85.Hd, 11.55.-m, 11.55.Bq}
\preprint{TAUP\,\,2939/11  \\

\today}

\begin{document}


\section{Introduction}

High energy QCD has reached a mature stage of development,
 in its description of  dilute-dense scattering.
Deep inelastic scattering with nuclei provides a good example of this.
 The main physical phenomena that emerges for this type of scattering has been discussed,
 and the non-linear equations that govern such processes have been derived and discussed in detail \cite{GLR,MUQI,MV,MUCD,B,K,JIMWLK,MBK}.
On the other hand, the scattering of the dense system of partons with the dense system of
 partons has been actively studied \cite{BRA,KOLU,LELU,BIIT,HIMST,MMSW,LMP}, but with limited success.
This is in spite of the fact that this scattering is closely
related to nucleus-nucleus scattering, which is the source of most of the experimental information, for the dense parton system.

At the moment there exist two general approaches to high energy QCD: the BFKL Pomeron calculus \cite{GLR,MUQI,BRA, BFKL,RevLI}, and the
Colour Glass Condensate approach (CGC) \cite{MV,JIMWLK}, which lead to the same non-linear equations \cite{B,K} for dilute-dense scattering.
The interrelation between these two approaches is not clear at the moment.
However, the equations that describe nucleus-nucleus collisions have not been derived,
in spite of considerable progress made in this direction
\cite{KOLU,BIIT,HIMST},  while in the BFKL Pomeron calculus, such equations have been proposed in Ref.\cite{BRA}.
These equations have been on the market for some time, but unfortunately, only three attempts to solve them are available in Refs.\cite{KLM,BOMO,BOBR}.

The main goal of this paper is to find the solution to these equations.
 In the next section we re-derive the equation of Ref.\cite{BRA} in momentum representation, which turn
out to be the most economical way
of finding the solution. In section 3 we will find the semi-classical solutions to the equation,
 which describe dense-dense scattering outside of the saturation region.
 Section 4 is devoted to finding the solution inside of the saturation region.
In the conclusion section we summarize our results.

\section{The BFKL Pomeron Calculus}
The goal of this section is to find the equation for nucleus-nucleus scattering  in the momentum representation
based on the BFKL Pomeron calculus, based on the main idea of Ref.\cite{BRA} that the equation for nucleus-nucleus collisions can be found from the equation of motion  for Pomerons.


\subsection{Functional integral formulation of the BFKL Pomeron Calculus $S_{0}$ }
The BFKL Pomeron calculus can be written through the functional integral \cite{BRA}
\begin{equation} \label{BFKLFI}
Z[\Phi, \Phi^+]\,\,=\,\,\int \,\,D \Phi\,D\Phi^+\,e^S \,\,\,\hspace{0.5cm}\mbox{with}\hspace{0.5cm}\,S \,=\,S_0
\,+\,S_I\,+\,S_E
\end{equation}
where $S_0$ describes free Pomerons, $S_I$ corresponds to their mutual interaction
while $S_E$ relates to the interaction with the external sources (target and
projectile). Here the free action is given by:
\bea
&&S_0= \int dY\prm\int d^2x_1\,d^2x_2
\Phi^\dag\Lb x_1,x_2,Y\prm\Rb  \nabla^{2}_{1}\nabla^{2}_{2}[ \frac{\partial}{\partial Y}+ H ] \Phi \Lb x_1,x_2,Y\prm\Rb
\eea

Define the following Fourier transform,
\bea
&&\Phi^\dag\Lb x_1,x_2,Y\prm\Rb\,\,\,\,\,\,=\,\,\,\,\,\,\Phi^\dag\Lb x_{12},b,Y\prm\Rb\,\,\,\,\,\, =\,\,\,\,\,\,x_{12}^2\int d^2k_1 e^{-ik_1\cdot x_{12}}\Phi^\dag\Lb k_1,b,Y\prm\Rb\lab{fourier}\\
&&\Phi\Lb x_1,x_2,Y\prm\Rb\,\,\,\,\,\,=\,\,\,\,\,\,\Phi\Lb x_{12},b,Y\prm\Rb\,\,\,\,\,\, =\,\,\,\,\,\,x_{12}^2\int d^2k_2 e^{ ik_2\cdot x_{12}}\Phi\Lb k_2,b,Y\prm\Rb\lab{fourier2}\eea

 Consider the free action as the sum of two terms:

 \bea\lab{S02P}
S_0\he S_0\prm + S_0\dpr\lab{somof2terms}
\eea
 where:
\bea
&&S_0\prm = \int dY\prm\int d^2x_1\,d^2x_2
\Phi^\dag\Lb x_1,x_2,Y\prm\Rb  \nabla^{2}_{1}\nabla^{2}_{2} \frac{\partial}{\partial Y}  \Phi \Lb x_1,x_2,Y\prm\Rb
\eea

This can be re-written as:

\bea
&&S\prm_0= 4\int dY\prm\int d^2b\,d^2x_{12}\int d^2k_1\,d^2k_2  \,e^{ ik_1\cdot x_{12}}
\Phi^\dag\Lb k_1,b,Y\prm\Rb x^{2}_{12} \nabla^2_1\nabla^2_2 \Lb x^2_{12}
\, e^{-ik_2\cdot x_{12}}\f{\D}{\D Y}  \Phi \Lb k_2,b,Y\prm\Rb\,\Rb
\hspace{0.8cm}\lab{SOI}\eea

where in the last step the replacement $d^2x_1\,d^2x_2 = 4\, d^2b \,d^2x_{12}$, where,

\bea
&&b=\f{x_1+x_2}{2}\hspace{3cm}
x_{12}=x_1-x_2
\lab{definitionb}\eea

where $b$
is the the impact parameter. In two-dimensional polar coordinates:

\bea
&&\nabla_{k_2}^2=\f{\D^2}{\D\,{k_2}^2}+\f{1}{k_2}\f{\D}{\D\,k_2}+\f{1}{k_2^2}\,\f{\D}{\D\,\theta}\lab{2dlaplacian}\eea

Let us introduce the next  change of variable  $ l=\ln k_2^2$,
such that the  two-dimensional Laplacian $ \nabla_{k_2}^2$ with respect to $k_2$,
simplifies to the following differential operator:
\bea
&&\nabla^2_{k_2}\he 4 e^{-l_{k_2}} \frac{\D^2}{\D l^2}+e^{-l_{k_2}}\f{\D}{\D\theta} \lab{DOP0}
\eea

According to the definition of \eq{DOP0}, the next term $ x_{12}^2\nabla_{x_1}^2\nabla_{x_2}^2(x_{12}^2 \,\,e^{-ik_1\cdot x_{12}})$ can be recast as,

\bea \lab{DOP1}
&&x_{12}^2\nabla_{x_1}^2\nabla_{x_2}^2\Lb x_{12}^2 \,\,e^{-ik_2\cdot x_{12}}\Rb\he
-x_{12}^2\nabla_{k_2}^2\nabla_{x_{12}}^4 e^{-ik_2\cdot x_{12}}\nn\\
\nn\\
&&\He -x_{12}^2\nabla_{k_2}^2\Lb k_2^4 \,\, e^{-ik_2\cdot x_{12}}\Rb\nn\\
\nn\\
&&\He\nabla_{k_2}^2\Lb k_2^4\,\,\nabla_{k_2}^2\,e^{-ik_2\cdot x_{12}}\,\Rb\nn\\
&&\He 16\Lb\,\Lb \f{\D}{\D l} +1\Rb^2+\f{\D}{\D\theta}\Rb\,\Lb\, \f{\D^2}{\D l^2}+\f{\D}{\D\theta}   \,\Rb  e^{-ik_2\cdot x_{12}}
\eea

where $\nabla^2_{k_2}$ is the two-dimensional Laplacian derivative with respect to $k_2$. Inserting \eq{DOP1}  back into \eq{SOI}, leads to the expression:

 \bea
&&S\prm_0= 4\int \! dY\prm\!\int \!d^2b\,d^2x_{12}\!\int\! d^2k_1d^2k_2\lab{SI4}\\
&&\times  \,e^{ ik_1\cdot x_{12}}
\Phi^\dag\Lb k_1,b,Y\prm\Rb \left\{ \nabla_{k_2}^2\Lb k_2^4\,\,\nabla_{k_2}^2\,e^{-ik_2\cdot x_{12}}\,\Rb\right\}\f{\D}{\D Y}  \Phi \Lb k_2,b,Y\prm\Rb\,
\hspace{1cm}\nn\\
\nn\\\nn\\
&&=32\int \! dY\prm\!\int \!d^2b\,d^2x_{12}\!\int\! d^2k_1d\theta dl\, e^l\lab{SI4a}\\
&&\times  \,e^{ ik_1\cdot x_{12}}
\Phi^\dag\Lb k_1,b,Y\prm\Rb \left\{ \Lb\,\Lb \f{\D}{\D l} +1\Rb^2+\f{\D}{\D\theta}\Rb\,\Lb\, \f{\D^2}{\D l^2}+\f{\D}{\D\theta}   \,\Rb  e^{-ik_2\cdot x_{12}}
\right\}\f{\D}{\D Y}  \Phi \Lb k_2,b,Y\prm\Rb\,
\hspace{1cm}\nn\eea

where in the last step, the  new  variable  $l$ was introduced. We use the convention
that derivatives only act on terms inside  of the curly brackets $\left\{\right\}$. Hence in \eq{SI4a}
derivatives only act on $e^{-ik_2\cdot x_{12}}$.
Assuming that
the $\theta$-dependent part of the integrand is a purely periodic function of $\theta$, then the derivatives can be re-ordered through
integration by parts, to yield:

 \bea
&&S\prm_0\he32\int \! dY\prm\!\int \!d^2b\,d^2x_{12}\!\int\! d^2k_1d\theta dl\,e^l \,e^{ ik_1\cdot x_{12}}\,e^{- ik_2\cdot x_{12}}\lab{SI4b}\\
&&\times  \,
\Phi^\dag\Lb k_1,b,Y\prm\Rb  \left\{ \Lb\,\Lb \f{\D}{\D l} +1\Rb^2+\f{\D}{\D\theta}\Rb\,\Lb\, \f{\D^2}{\D l^2}+\f{\D}{\D\theta}   \,\Rb
\f{\D}{\D Y}  \Phi \Lb k_2,b,Y\prm\Rb\,\right\}
\hspace{1cm}\nn\\
\nn\\\nn\\
&&\He
32\int \! dY\prm\!\int \!d^2b\,d^2x_{12}\!\int\! d^2k_1d\theta dl\,e^l \,e^{ ik_1\cdot x_{12}}\,e^{- ik_2\cdot x_{12}}\lab{SI4c}\\
&&\times  \,
\Phi^\dag\Lb k_1,b,Y\prm\Rb  \left\{ \,\Lb \f{\D}{\D l} +1\Rb^2\,\, \f{\D^2}{\D l^2}   \,\,
\f{\D}{\D Y}  \Phi \Lb k_2,b,Y\prm\Rb\,\right\}
\hspace{1cm}\nn\eea

where in the last step it was assumed that $\Phi \Lb k_2,b,Y\prm\Rb$ is purely a function of $ k_2^2 =e^l$ and not a function of the angular
coordinate $\theta$, such that $\theta$-derivatives vanish. Now returning to the integration variables $\int d\theta dl\,e^l = 2\int d^2k_2$
where $l=\ln k_2^2$, and integrating over $x_{12}$ leads to the delta function $\Lb 2\pi\Rb^2\de^2\Lb k_1-k_2\Rb$. The delta function is absorbed by the
 integral over $k_2$-space resulting in the expression:
 
\bea
&&S\prm_0\he
64\Lb 2\pi\Rb^2\int \! dY\prm\!\int \!d^2b\,\!\int\! d^2k_1\,\Phi^\dag\Lb k_1,b,Y\prm\Rb  \left\{ \,\Lb \f{\D}{\D l} +1\Rb^2\,\, \f{\D^2}{\D l^2}   \,\,
\f{\D}{\D Y}  \Phi \Lb k_1,b,Y\prm\Rb\,\right\}
\hspace{1cm}\lab{SI4d}\eea

where $l=\ln k_1^2$. This  part of the action gives the following contribution  to the  equation of motion, which stems from the condition
 $\de S_0/\de\Phi^\dag(k,b,Y)=0$:

\bea  \label{SI6}
&&\delta S\prm_0/\de\Phi^\dag(k,b,Y)=  64 (2\pi)^{2}  \Lb \f{\D}{\D l} +1\Rb^2\,\, \f{\D^2}{\D l^2}   \,\,
\f{\D}{\D Y}  \Phi \Lb k,b,Y\prm\Rb\,
\eea

Recall that in the above calculation,
 for the sake of simplicity we considered $S_0=S_0\prm +S_0\dpr$, and we
calculated the Fourier transform of the $S_0\prm$ part.
  The full variation $\de S_0/\delta\Phi^\dag(k,b,Y)=\de \Lb S_0\prm+S_0\dpr\Rb/\de\Phi^\dag\Lb k,b,Y\Rb\,\,$ 
is given by the expression:

\bea \label{SI8}
&&\delta S_0/\delta\Phi^\dag(k,b,Y)\he 64 (2\pi)^2
 \Lb \f{\D}{\D l} +1\Rb^2\,\, \f{\D^2}{\D l^2}   \,\, \Lb
\f{\D}{\D Y} -H\Rb\Phi \Lb k,b,Y\prm\Rb
\eea

The general properties of the Hamiltonian $H$ have been discussed in Refs. \cite{RevLI, BRA} in coordinate representation,  which read as follows in momentum space representation:
\bea
&&
H f(k,Y) \,\,= \,\,\frac{\bas}{2
\pi}\,\int\,d^2 l\, K\Lb k,l\Rb\,\left\{
f(k,l,Y)\,-\,f(k,Y) \right\} \hspace{1cm} \label{H}\\
\nn\\
&& K\Lb k,l\Rb\,\,=\,\,\frac{
k^2_\perp}{l^2_\perp\,(\vec{k} - \vec{l})^2_\perp}
\label{K}\eea

\begin{boldmath}
\subsection{Interaction  term of the action $S_{I}$ and the  Field equation}
\end{boldmath}
The mutual interaction of the Pomerons is described  by $S_{I}$.  The interaction which is related
to the  external sources (target and projectile) are not consider in this paper.
In this approach $S_{I}$  is given by:

\bea
S_I&&=\f{2\pi\bas^2}{N_c}\int dY\prm\int\f{d^2x_1\,d^2x_2\,d^2x_3}{x_{12}^2x_{23}^2x_{31}^2}\left\{\Lb L_{12}\Phi\Lb x_1,x_2,Y\prm\Rb\Rb
\Phi^\dag\Lb x_2,x_3,Y\prm\Rb\Phi^\dag\Lb x_3,x_1,Y\prm\Rb\right.\hspace{0.6cm}\lab{SI}\\&&\left. + \Lb L_{12}\Phi^\dag\Lb x_1,x_2,Y\prm\Rb\Rb
\Phi\Lb x_2,x_3,Y\prm\Rb\Phi\Lb x_3,x_1,Y\prm\Rb \right\}\nn\\
\nn\\
&&=\f{2\pi\bas^2}{N_c}\,\int dY\prm\int\f{\,4d^2b\,d^2x_{12}\,d^2x_3}{x_{12}^2x_{23}^2x_{31}^2}\left\{\Lb L_{12}\Phi\Lb x_1,x_2,Y\prm\Rb\Rb
\Phi^\dag\Lb x_2,x_3,Y\prm\Rb\Phi^\dag\Lb x_3,x_1,Y\prm\Rb\right.\hspace{0.6cm}\lab{SIb}\\&&\left. + \Lb L_{12}\Phi^\dag\Lb x_1,x_2,Y\prm\Rb\Rb
\Phi\Lb x_2,x_3,Y\prm\Rb\Phi\Lb x_3,x_1,Y\prm\Rb \right\}\nn\eea

where in the last step the replacement $d^2x_1  d^2x_2 =4 d^2b d^2x_{12}$, (see \eq{definitionb}), and where the following
differential operator was introduced:

\bea
&&L_{12}=x_{12}^4\nabla_{x_1}^2\nabla_{x_2}^2 \lab{L12b}
\eea

Now the above defined Fourier transform of Eqs. (\ref{fourier}) and (\ref{fourier2}) are inserted into \eq{SIb}, where 
its assumed that  $b\gg x_{12}$,
where $b=\Lb x_1+x_2\Rb/2$, i.e. the impact parameter
is much bigger than the size of the dipole. Inserting the Fourier transforms of Eqs. (\ref{fourier}) and (\ref{fourier2}) into \eq{SIb} gives,

\bea \lab{zeta}
&&S_I=\f{2\pi\bas^2}{N_c}\,\int dY\prm\int\,4\,d^2b\,d^2x_{12}\,d^2x_3\int\, d^2k_1\, d^2k_2\, d^2k_3\,
\, \lab{SI21}\\
&&  \left\{ \exp\Lb ik_2\cdot x_{23} + ik_3\cdot x_{31} \Rb\Lb \ml{ \f{ L_{12}\Lb x_{12}^2\, e^{-ik_1\cdot x_{12}}\Rb}{x_{12}^2}}\Rb\,\,\,\,\,\,
\Phi\Lb k_1,b,Y\prm\Rb\Phi^\dag\Lb k_2,b,Y\prm\Rb\Phi^\dag\Lb k_3,b,Y\prm\Rb
\right. \nn\\
\nn\\
&& \left. + \,\,\,  \exp\Lb - ik_2\cdot x_{23} - ik_3\cdot x_{31} \Rb
\Lb \ml{\f{ L_{12}\Lb x_{12}^2\, e^{ ik_1\cdot x_{12}}\Rb}{x_{12}^2}}\Rb\,\,\,\,\,\, \Phi^\dag\Lb k_1,b,Y\prm\Rb\Phi\Lb k_2,b,Y\prm\Rb\Phi\Lb k_3,b,Y\prm\Rb \right\} \nn
\nn
\eea

The integration over $x_3$ leads to the Dirac delta function $\Lb 2\pi\Rb^2\,\de^2\Lb k_2-k_3\Rb$, where $\de^2\Lb k_2-k_3\Rb$ labels
the delta function in two dimensions. The delta function is absorbed by the integration over $k_3$ which leads to:

\bea
&&S_I=\f{2\pi\bas^2}{N_c} \,\,(2\pi)^2 \,\int dY\prm\int\,4\,d^2b\,d^2x_{12}\,\int\, d^2k_1\, d^2k_2\,
\, \lab{SI20}\\
&&  \left\{ \exp\Lb -ik_2\cdot x_{12}  \Rb\Lb \ml{ \f{ L_{12}\Lb x_{12}^2\, e^{-ik_1\cdot x_{12}}\Rb}{x_{12}^2}}\Rb\,\,\,\,\,\,
\Phi\Lb k_1,b,Y\prm\Rb\Phi^\dag\Lb k_2,b,Y\prm\Rb\Phi^\dag\Lb k_2,b,Y\prm\Rb
\right. \nn\\
\nn\\
&& \left. + \,\,\,  \exp\Lb  ik_2\cdot x_{12} \Rb
\Lb \ml{\f{ L_{12}\Lb x_{12}^2\, e^{ ik_1\cdot x_{12}}\Rb}{x_{12}^2}}\Rb\,\,\,\,\,\, \Phi^\dag\Lb k_1,b,Y\prm\Rb\Phi\Lb k_2,b,Y\prm\Rb\Phi\Lb k_2,b,Y\prm\Rb \right\} \nn
\nn
\eea

According to the definition of \eq{L12b}, the term $L_{12}\Lb x_{12}^2 \exp\Lb-ik_1\cdot x_{12}\Rb\Rb\,/\,x_{12}^2$ can be recast as,

\bea\lab{L12recast1}
&&\ml{\f{L_{12}\Lb x_{12}^2 e^{-ik_1\cdot x_{12}}\Rb}{x_{12}^2}} \,\,\,=\,\,\,x_{12}^2\nabla_{x_1}^2\nabla_{x_2}^2\Lb x_{12}^2 \,\,e^{-ik_1\cdot x_{12}}\Rb\nn\\
\\&&=
 \nabla_{k_1}^2\Lb k_1^4\,\,\nabla_{k_1}^2\,e^{-ik_1\cdot x_{12}}\,\Rb
\nn\\&&=  16\Lb\,\Lb \f{\D}{\D l} +1\Rb^2+\f{\D}{\D\theta}\Rb\,\Lb\, \f{\D^2}{\D l^2}+\f{\D}{\D\theta}   \,\Rb  e^{-ik_1\cdot x_{12}}
\eea

where $l=\ln k_1^2$ and where the last expression was discussed  and introduced in the first section (see \eq{DOP1}). Inserting this result into the interaction action of \eq{SI20},
 after some algebra one arrives at the expression:

 \bea&&S_I\he 16 \,\, \f{2\pi\bas^2}{N_c} \,\,(2\pi)^2 \int dY\prm\int\,4\,d^2b\,d^2x_{12}\, \int\, d^2k_1\, d^2k_2\,\lab{SI2}\\
&&\left\{\,\exp\Lb -i\Lb k_1+k_2\Rb\cdot x_{12}\Rb \,\Lb \f{\D}{\D l} +1\Rb^2\,\f{\D^2}{\D l^2}
\Phi\Lb k_1,b,Y\prm\Rb\Phi^\dag\Lb k_2,b,Y\prm\Rb\Phi^\dag\Lb k_2,b,Y\prm\Rb\right.\nn\\
&&\left. +\,\, \exp\Lb i\Lb k_1+k_2\Rb\cdot x_{12}\Rb \Lb \f{\D}{\D l} +1\Rb^2\,\f{\D^2}{\D l^2}
\Phi^\dag\Lb k_1,b,Y\prm\Rb\Phi\Lb k_2,b,Y\prm\Rb\Phi\Lb k_2,b,Y\prm\Rb\right\} \nn
\eea

 where it was assumed that the functions $\Phi\Lb k_1,b,Y\prm\Rb $ and $\Phi^\dag\Lb k_1,b,Y\prm\Rb$
are purely functions of the radial $k_1^2$ coordinate and do not depend on the angular coordinate $\theta$, such that $\theta$ derivatives that appeared
 in \eq{L12recast1}
have been dropped. The integration over $x_{12}$ leads to the Dirac delta function $\Lb 2\pi\Rb^2\,\de^2\Lb k_l+k_2\Rb$.
 The delta function is absorbed by the integration over $k_2$ which leads to:

 \bea&&S_I\he 16 \,\, \f{2\pi\bas^2}{N_c} \,\,(2\pi)^4\int dY\prm\int\,4\,d^2b\,\, \int\, d^2k_1\, \,\lab{SFF1}\\
&&\left\{\,\Phi^\dag\Lb -k_1,b,Y\prm\Rb\Phi^\dag\Lb - k_1,b,Y\prm\Rb\,\Lb \f{\D}{\D l} +1\Rb^2\,\f{\D^2}{\D l^2}
\Phi\Lb k_1,b,Y\prm\Rb\right.\nn\\
&&\left.\Phi\Lb -k_1,b,Y\prm\Rb \Phi\Lb -k_1,b,Y\prm\Rb \Lb \f{\D}{\D l} +1\Rb^2\,\f{\D^2}{\D l^2}
\Phi^\dag\Lb k_1,b,Y\prm\Rb\right\} \nn
\eea

 where $l=\ln k_1^2$. Integrating the last term in \eq{SFF1} by parts,
 and taking into account that  $d^2 k_1 = \h d \theta\, d l\, e^{l}$ where $\theta$ is the azimuthal angle,
\eq{SFF1} simplifies to the following expression:
\bea&&S_I\he 16 \,\, \f{2\pi\bas^2}{N_c} \,\,(2\pi)^4\int dY\prm\int\,4\,d^2b\,\, \int\, d^2k_1\, \,\lab{SFF2}\\
&&\left\{\,\Phi^\dag\Lb -k_1,b,Y\prm\Rb\Phi^\dag\Lb - k_1,b,Y\prm\Rb\,\Lb \f{\D}{\D l} +1\Rb^2\,\f{\D^2}{\D l^2}
\Phi\Lb k_1,b,Y\prm\Rb\right.\nn\\
&&\left.\,\Phi^\dag\Lb k_1,b,Y\prm\Rb\Lb \f{\D}{\D l} +1\Rb^2\,\f{\D^2}{\D l^2} \Phi\Lb -k_1,b,Y\prm\Rb \Phi\Lb -k_1,b,Y\prm\Rb \,
\right\} \nn
\eea

The last equation \eq{SFF2} allows us to  study the classical equation of the action, which can be obtained from the conditions \cite{BRA,KLM}:
$\ml{\frac{\delta S}{\delta\Phi(k, b, Y)}}=0 $ and $\ml{\frac{\delta S}{\delta\Phi^\dag(k, b, Y)}}=0 $

\subsection{Classical equations }

The  functional derivative  of the action  for the  effective  pomeron field theory,
 with respect to $\Phi^\dag(k, b, Y)$    can be found from the results of Eqs. (\ref{SI8}) and  (\ref{SFF2}),
which give:

\bea &&
\delta \Lb S_0\,+\, S_I\Rb / \delta \Phi^{\dag}(k, b, Y)\he64 (2\pi)^{2}  \,\,\Lb\frac{\partial}{\partial l}\,+\,1\Rb^2\frac{\partial^{2}}{\partial l^{2}} \, \Big(\frac{\partial}{\partial Y} \,\,-\,\,H\Big) \Phi \Lb k,b,Y \,\Rb\,\nn\\
&&+\, 16\Lb\f{2\pi\bas^2}{N_c}\Rb\Lb 2\pi\Rb^4 \left\{  2\Phi^{\dag}\Lb -k,b,Y\prm\Rb  \Lb\frac{\partial}{\partial l}+1\Rb^2
\f{\partial^2}{\partial l^2} \Phi\Lb k,b,Y\prm\Rb +  \Lb\frac{\partial}{\partial l}+ 1\Rb^2\f{\partial^2}{\partial l^2} \Phi^2 \Lb -k,b,Y\prm\Rb\right\}
\hspace{0.7cm}\lab{CEQ10}
\eea

 where $l=\ln k^2$. The equation for nucleus-nucleus scattering can be derived from the following equation of motion\cite{BRA}:
 
 \bea \label{EQOM1}
&& \delta \Lb\, S_0\,+\, S_I\Rb\, / \delta \Phi^\dag\Lb\,k, b, Y\Rb\,\,\,=\,\, 0\,\,\,\,\mbox{and}\,\,\,
  \delta \Lb\, S_0\,+\, S_I\Rb\, / \delta \Phi\Lb\,k, b, Y\Rb\,\,\,=\,\, 0
  \eea

Now the following approach is used to average these equations: 

 \beq \label{AVRG}
\Big{ \langle} {\cal O} \Lb k, Y; b \Rb \Big{\rangle}\,\,\,=\,\,\,\frac{\int  D \Phi D \Phi^\dag\,{\cal O}\Lb k, Y; b\Rb\,e^{S\Lb \Phi,\Phi^\dag\Rb}}{\int  D \Phi D \Phi^\dag\,e^{S\Lb \Phi,\Phi^\dag\Rb }}
\eeq

Introducing the following new functions:

\beq \label{AMPL}
N\Lb k,Y;b\Rb \he \,2 \pi^2 \as  \Big{\langle} \Phi\Lb k, Y;b \Rb\Big{\rangle}
\hspace{2cm}
  N^\dag\Lb k,Y;b\Rb \he \, 2 \pi^2 \as  \Big{\langle} \Phi^\dag\Lb k, Y;b \Rb\Big{\rangle}
  \eeq

then the equation of motion of \eq{CEQ10} reduces to:

 \bea \label{EQOM1}
&&0\,\,\,=\,\,\Lb\,\frac{\partial}{\partial l}\,+\,1\Rb^{2}\,\frac{\partial^{2}}{\partial l^{2}}\, \, \Lb\,\frac{\partial}{\partial Y'} \,\,-\,\,H\Rb\, N \Lb\, k_l,b,Y'\,\Rb\,\,\,\\
&& \,\,+\,\bas\left\{ 2\,N^\dag\Lb\, -k_l,b,Y\prm\Rb\,\Lb\,\frac{\partial}{\partial l}\,+\,1\Rb^{2}\,\frac{\partial^{2}}{\partial l^{2}}\, \,\,
N\Lb\, k_l,b,Y\prm\Rb\,\,\,+\,\, \Lb\,\frac{\partial}{\partial l}\,+\, 1\Rb^{2}\,\frac{\partial^{2}}{\partial l^{2}}\, N^2 \Lb\, -k_l,b,Y\prm\Rb\,\,\,\right\}
\nn
\eea

Using the following property of the BFKL Pomeron, namely,

\beq \label{BFKLH}
\,p^2_1\,p^2_2\,\Lb\,
\frac{\partial}{\partial Y} - H \Rb\, \,\,=\,\,\Lb\,
\frac{\partial}{\partial Y} - H^\dag \Rb\,\,p^2_1\,p^2_2,
\eeq

we can derive the second equation of motion, by taking  the functional derivative with respect to $\Phi$.
This generates the following equation of motion:

 \bea \label{EQOM2}
&&0\,\,\,=\,\,\Lb\,\frac{\partial}{\partial l}\,+\,1\Rb^{2}\,\frac{\partial^{2}}{\partial l^{2}}\, \, \Lb\,-\,\frac{\partial}{\partial Y} \,\,-\,\,H\Rb\, N^\dag \Lb\, k_l,b,Y \,\Rb\,\,\,\\
&& \,\,+\,\bas\left\{ 2\,N\Lb\, -k_l,b,Y\prm\Rb\,\Lb\,\frac{\partial}{\partial l}\,+\,1\Rb^{2}\,\frac{\partial^{2}}{\partial l^{2}}\, \,\,
N^\dag\Lb\, k_l,b,Y\prm\Rb\,\,\,+\,\, \Lb\,\frac{\partial}{\partial l}\,+\, 1\Rb^{2}\,\frac{\partial^{2}}{\partial l^{2}}\, \Lb\, N^\dag\Rb^2 \Lb\, -k_l,b,Y\prm\Rb\,\,\,\right\}
\nn\eea

In \eq{EQOM1} and \eq{EQOM2}  the following identities were implemented:

\bea 
  \Big{\langle} \Phi^2\Lb k, Y;b \Rb\Big{\rangle}&&\he \Lb \Big{\langle} \Phi\Lb k, Y;b \Rb\Big{\rangle}\Rb^2\lab{N2}\\
\Big{\langle} \Lb\Phi^\dag\Rb^2\Lb k, Y;b \Rb\Big{\rangle}&&\he \Lb \Big{\langle} \Phi^\dag\Lb k, Y;b \Rb\Big{\rangle}\Rb^2\nn \\
\Big{\langle} \Phi\Lb k, Y;b \Rb\,\Phi^\dag\Lb k, Y; b\Rb\Big{\rangle} &&\he \Big{\langle} \Phi\Lb k, Y;b \Rb\Big{\rangle}\times  \Big{\langle} \Phi^\dag\Lb k, Y;b \Rb\Big{\rangle}\nn
   \eea
For a discussion on  why these equations are correct, within an accuracy of about $1/A^{1/3}$ in the case of nucleus-nucleus scattering,
see Refs.\cite{BRA,KLM}.

\section{Semiclassical solution}
\subsection{The system of equations: general approach}
In this section we find the semiclassical solutions to \eq{EQOM1} and \eq{EQOM2}. In the semi-classical approximation, we are searching  for  solutions of the following form \cite{BKL}:
\bea
N\Lb l\equiv \ln(k^2), b,Y^{\prime}\Rb
\,\,&=&\,\exp\Lb S\Lb l , b,Y^{\prime}\Rb\Rb \hspace{1.25cm}\mbox{where}\hspace{1cm}S\,=\,\om\,Y^{\prime}\,-\,( 1 \,\,-\,\,\gamma)\,l\,+\,\beta(b)\nn
\\
N^\dag\Lb l\equiv \ln(k^2), b,Y^{\prime}\Rb
\,\,&=&\,\exp\Lb  S^\dag\Lb l, b,Y^{\prime}\Rb\Rb \hspace{1cm}\mbox{where}\hspace{1cm}S^\dag\,=\,\om^\dag\,Y^{\prime}\,-\,(1\,\,-\,\,\gamma^\dag)\,l\,+\,\beta^\dag(b)
\hspace{1cm}\lab{SC1}\eea
where $\omega(l ,b,Y^{\prime})$\, and \, $\gamma(l ,b,Y^{\prime})$ are smooth functions of
$Y^{\prime}$ and $l$, and the following conditions are assumed:
\bea
&& \f{d \omega(l ,b,Y^{\prime})}{d Y^{\prime}}\hspace{0.3cm}\ll\hspace{0.3cm}\omega^2(l ,b,Y^{\prime});\,\,\,\,\f{d \omega(l ,b,Y^{\prime})}{d l}
\hspace{0.3cm}\ll\hspace{0.3cm}\omega(l ,b,Y^{\prime})\,\Lb 1 - \gamma(l ,b,Y^{\prime})\Rb;\,\,\, \lab{cond2}\\
\nn\\
&&\f{d
\gamma(l ,b,Y^{\prime})}{d
l} \hspace{0.3cm}\ll\hspace{0.3cm}\Lb 1 - \gamma(l ,b,Y^{\prime})\Rb^2;\,\,\,\,\,\f{d
\gamma(l ,b,Y^{\prime})}{d Y^{\prime}}
\hspace{0.3cm}\ll\hspace{0.3cm}\omega(l ,b,Y^{\prime})\,\Lb 1 - \gamma(l ,b,Y^{\prime})\Rb;\lab{cond3}\eea
 with analogous conditions for the functions $\om^\dag(l ,b,Y^{\prime})$ and $\ga^\dag(l ,b,Y^{\prime})$.
Assuming that for \eq{SC1},  the method of characteristics  can be applied (see, for
example, Ref. \cite{MATH}) to solve the non-linear equation.  Notice that
\bea \label{HSC}
&&H e^{S}\he\bas \int K\Lb k,k'\Rb e^{S\Lb k',b,Y\Rb}\,\,=\bas \chi(\gamma)  e^{S\Lb k,b,Y\Rb}\\
\nn\\
\mbox{where}\hspace{0.5cm} &&\chi\Lb \ga\Rb\he 2 \psi(1) \,\,-\,\,\psi\Lb \ga
\Rb \,\,-\,\,\psi\Lb 1 \,-\,\ga\Rb\lab{chi}
\eea
where   $\psi(z)\,=d \ln \Ga(z)/d z$ is the Di-Gamma function. In the semiclassical approach (for $\ga$  a smooth function), inserting the definition of \eq{SC1} into the equations
 of motion of
 \eq{EQOM1} and \eq{EQOM2}, and using the conditions of Eqs.(\ref{cond2}) - (\ref{cond3}),
leads to the following formulae in the semi-classical approach:
\bea \label{EQSC}
&&\om\,\,-\,\,\bas \chi\Lb \gamma\Rb \,\,+\,\,\kappa\Lb \gamma^\dag\Rb\bas\,e^{\tilde{S}^\dag}\,\,+\,\,\bas\,e^{\tilde{S}}\hspace{0.8cm}\He 0\nn\\
&&-\,\om^\dag\,\,-\,\,\bas \chi\Lb \gamma^\dag\Rb \,\,+\,\,\kappa\Lb \ga\Rb\bas\,e^{\tilde{S}^\dag}\,\,+\,\,\bas\,e^{\tilde{S}}\he 0\eea
where the following functions were introduced with definitions:
\bea
&&\kappa\Lb \ga\Rb\he\frac{2\,\ga^2}{4 (2 \ga - 1)^2}\hspace{1.5cm}\tilde{S}\he S\,-\ln \kappa(\ga)
\hspace{1.5cm}\tilde{S}^\dag\he S^\dag\,-\ln \kappa(\ga^\dag)\lab{definitions}\eea

For the equation in the form
\beq \label{SC2}
F(Y^{\prime},l,\tilde{S},\gamma,\omega)=0
\eeq
where $S$ is given by \eq{SC1}, we can introduce the set
of
characteristic lines on which  $l(t), Y^{\prime}(t), S(t),$ $ \omega(t),$ and $
\gamma(t)$ are
functions of the variable $t$ (which we call artificial time), that satisfy the following
equations:
\begin{eqnarray}
&&(1.)\,\,\,\,\,\,\,\frac{d l}{d\,t}\,\,=\,\,F_{\gamma}\,\,=\,-\,\bas\,\frac{d \chi(\ga)}{d \ga}\nn\\
\,\,\,\,\,\,\,\,\,\,\,\,&&(2.)\,\,\,\,\,\,\,
\frac{d\,Y^\prime}{d\,t}\,\,=\,\,F_{\omega}\,\,=\,\,1\,\,\,\,\,\,\,\nn\\
 &&(3.)\,\,\,\,\,\,\,
\frac{d\,\tilde{S}}{d\,t}\,\,=\,\,\gamma\,F_{\gamma}\,+\,\omega\,F_{\omega}\,\,=\,\,  \bas\,\Lb 1 \,-\,\ga\Rb\,\frac{d \chi(\ga)}{d \ga}\,\,+\,\,\om\nonumber \\
&&(4.)\,\,\,\,\,\,\,\frac{d\,\gamma}{d\,t}\,\,=\,\,-
(\,F_{l}\,+\,\gamma\,F_{S}\,)\,\,=\,\,\,\bas \kappa\Lb \ga^+\Rb\,(1 - \ga^\dag)\,e^{\tilde{S}^\dag}\,\,+\,\,\bas\,(1\,-\,\ga)\,e^{\tilde{S}}\nn\\
&& (5.)\,\,\,\,\,\,\,
\frac{d\,\omega}{d\,t}\,\,=\,\,- \,(\,F_{Y^{\prime}}  \,+\,\omega\,F_S\,)\,\,=\,\,\,-\,\bas\,\om^\dag
\kappa\Lb \ga^+\Rb\,e^{\tilde{S}^\dag}\,\,-\,\,\bas\,\om\,e^{\tilde{S}}
\label{SCEQ1}
\end{eqnarray}
where $F_l =\ml{ \frac{\partial F(Y^{\prime},l,\tilde{S},\gamma,\omega)}{\partial l}}$, and the terms in the first of Eqns. (\ref{EQSC})
 that depend on $\tilde{S}^\dag,\om^\dag\,\ga^\dag$,
are treated as if they depend explicitly  on $Y^\prime$ and $l$. The same five equations we can write for the second of Eqns. (\ref{EQSC}) which have the following form:
\begin{eqnarray}
&&(1.)\,\,\,\,\,\,\,\frac{d l^\dag}{d\,t}\,\,=\,\,F_{\gamma^\dag}\,\,=\,-\,\bas\,\frac{d \chi(\ga^\dag)}{d \ga^\dag}\nn\\
\,\,\,\,\,\,\,\,\,\,\,\,&&(2.)\,\,\,\,\,\,\,
\frac{d\,Y^{\prime,\dag}}{d\,t}\,\,=\,\,F_{\omega^\dag}\,\,=\,\,-1,\,\,\,\,\,\,\,\nn\\
 &&(3.)\,\,\,\,\,\,\,
\frac{d\,\tilde{S}^\dag}{d\,t}\,\,=\,\,\gamma^\dag\,F_{\gamma\dag}\,+\,\omega^\dag\,F_{\omega^\dag}\,\,=\,\,  \bas\,\Lb 1\,-\,\ga^\dag\Rb\,\frac{d \chi(\ga^\dag)}{d \ga^\dag}\,\,-\,\,\om^\dag\nonumber \\
&&(4.)\,\,\,\,\,\,\,\frac{d\,\gamma^\dag}{d\,t}\,\,=\,\,-
(\,F_{l}\,+\,\gamma^\dag\,F_{\tilde{S}}\,)\,\,=\,\,\bas \kappa\Lb \ga\Rb\,\Lb 1\,-\,\ga\Rb\,e^{\tilde{S}}\,\,+\,\,\bas\,\Lb1\,-\,\ga^\dag\Rb\,
e^{\tilde{S}^\dag}\nn\\
&& (5.)\,\,\,\,\,\,\,
\frac{d\,\omega^\dag}{d\,t}\,\,=\,\,- \,(\,F_{Y^{\prime}}  \,+\,\omega^\dag\,F_S\,)\,\,=\,\,\,\,-\,\bas\,\om
\kappa\Lb \ga\Rb\,e^{\tilde{S}}\,\,\,-\,\bas\,\om^\dag\,e^{\tilde{S}^\dag}
\label{SCEQ2}
\end{eqnarray}
\subsection{The system of equations: linear approximation}
We start with the solution of \eq{EQOM1} and \eq{EQOM2} in the kinematic region where the non-linear corrections are small.
In this region, we can reduce this system of equations to the solution of the BFKL equation,
 both for $N\Lb l , b,Y^{\prime}\Rb$ and $N^\dag\Lb l , b,Y^{\prime}\Rb$. Actually
in order to find the solution,
it isn't necessary to solve the equation, since  the $t$-channel unitarity constraints for the BFKL Pomeron can be used instead (see Refs. \cite{BFKL,GLR,MUSH}).
This is given by:
\beq \label{BFKLUC}
N_{BFKL}\Lb L , b,Y\Rb\,\,\,=\,\,\,\int d^2 b' \int d l \,N_{BFKL}^\dag\Lb L - l  , \vec{b}- \vec{b}', Y - Y^{\prime}\Rb\,\,N_{BFKL}\Lb l , b',Y^{\prime}\Rb
\eeq

Using the explicit form for the  BFKL solution \cite{RevLI,BFKL}, then \eq{BFKLUC} reduces to the following expression:
\bea \label{BFKLUC1}
&&N_{BFKL}\Lb L , b,Y\Rb\,\,\, =\,\,\,\int \,\frac{d \ga}{2 \pi i}\,n_{BFKL}\Lb \ga, b \Rb \,e^{\bas \chi(\ga)\,Y\,\,-\,\,( 1 - \ga)\,L}\,\,\\
&&=\,\,\int d^2 b' \int d \,l\int \,\frac{d \ga}{2 \pi i}\,\int \,\frac{d \ga'}{2 \pi i}
\,n^\dag_{BFKL}(\ga, \vec{b}-\vec{b}')\,e^{ \bas \chi(\ga)\,\Lb Y\,-\,Y^{\prime}\Rb\,\,-\,\,( 1 - \ga)\,\Lb L\,-\,l\Rb}\,n_{BFKL}\Lb \ga', b'\Rb e^{ \bas \chi(\ga')\,\,Y^{\prime}\,\,-\,\,(1 - \ga')\,l}\nn
\eea
where $L\,\,=\,\,\ln\Lb k^2_p/k^2_t\Rb$ and $ l\,\,=\,\,\ln\Lb k^2/k^2_t\Rb$.  $k_p$ and $k_t$ are the momenta of the dipoles in the projectile and the target, respectively.
 The function $n_{BFKL}\Lb \ga, b \Rb $ is determined by the initial condition at $Y = 0$.

The integration over $l$ leads to the delta function $\Lb 2\pi\Rb\de\Lb\ga-\ga\prm\Rb$, and hence the $\ga\prm$ integral
is immediately solvable by setting $\ga=\ga\prm$ everywhere in the integrand. Integrating over the impact parameter $b\prm$ yields:
\bea
\int d^2 b' \,n^\dag_{BFKL}(\ga, \vec{b}-\vec{b}')
\,n^{BFKL}(\ga, b')\,\,=\,\,n_{BFKL}(\ga, b)\lab{bintegral}\eea

 In light of this \eq{BFKLUC1} simplifies to
\bea \label{BFKLUC1a}
&&N_{BFKL}\Lb L , b,Y\Rb\he \int \,\frac{d \ga}{2 \pi i}\,
\,\,e^{ \bas \chi(\ga)\, Y\,\,-\,\,(1 - \ga)\,L }\,n_{BFKL}\Lb \ga, b\Rb
\eea

Incidentally \eq{BFKLUC1a} satisfies \eq{BFKLUC}.
Therefore, we have reproduced \eq{BFKLUC}, where $N^\dag$ and $N$ are given explicitly by:
\bea
&&N_{BFKL}^\dag\Lb L - l  , \vec{b}- \vec{b}',Y - Y^{\prime}\Rb\,\,=\,\,\int \,\frac{d \ga}{2 \pi i}\,n^\dag_{BFKL}(\ga, \vec{b}-\vec{b}')\,e^{ \bas \chi(\ga)\,\Lb Y\,-\,Y^{\prime}\Rb\,\,-\,\,(1 - \ga)\,\Lb L\,-\,l\Rb}\label{BFKLUC2}\\
&&N_{BFKL}\Lb l , b',Y^{\prime}\Rb\,\,=\,\,\int\frac{d \ga'}{2 \pi i}\,n_{BFKL}\Lb \ga', b'\Rb e^{ \bas \chi(\ga')\,\,Y^{\prime}\,\,+\,\,( 1 - \ga')\,l}\label{BFKLUC3}
\eea
where the functions $n^\dag_{BFKL}(\ga, \vec{b}-\vec{b}')$ and $n_{BFKL}\Lb \ga', b'\Rb$ can be found from the initial conditions for
 $N_{BFKL}^\dag$ at $Y' = Y$ and  $N_{BFKL} $ at $Y'=0$.

\FIGURE[ht]{
\epsfig{file=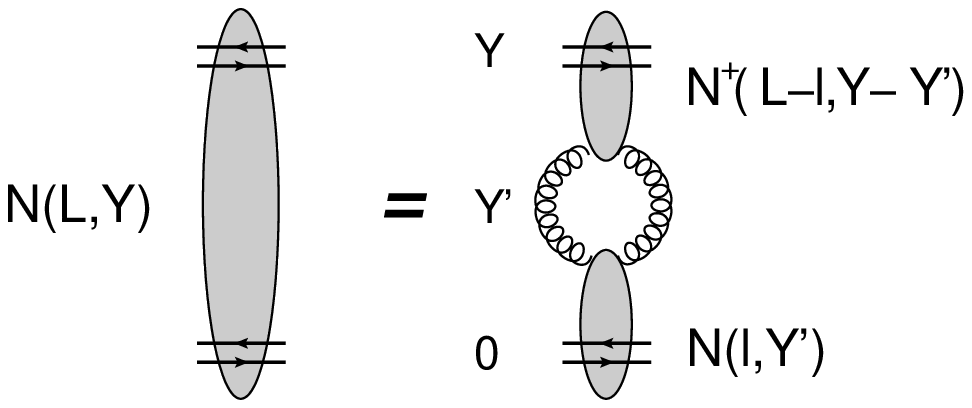,width=70mm}
\caption{The BFKL Pomeron: $t$-channel unitarity.}
\label{lineq}}
 The integral expressions of Eqs. (\ref{BFKLUC2}) and (\ref{BFKLUC3}) can be used to derive the explicit
expressions for \\
$N_{BFKL}^\dag\Lb L - l  , \vec{b}- \vec{b}',Y - Y^{\prime}\Rb$ and $N_{BFKL}\Lb l , b',Y^{\prime}\Rb$
in the semi-classical approach, by integrating  using the method of steepest decent.
This is assuming that
$n^\dag_{BFKL}(\ga, \vec{b}-\vec{b}')$ and $n_{BFKL}\Lb \ga', b'\Rb$ are  continuous, and differentiable functions of $\ga$ and $\ga'$.
\vspace{1cm}

 The saddle point equations, for the functions in the exponent in Eqs. (\ref{BFKLUC2}) and (\ref{BFKLUC3}) are:
\bea
&&\bas\f{d\chi\Lb\ga_{SP}\Rb}{d\ga_{SP}}\Lb Y-Y\prm\Rb +L-l\he 0\,\,\,\mbox{and}\,\,\,
\bas\f{d\chi\Lb\ga\prm_{SP}\Rb}{d\ga\prm_{SP}}Y\prm +l\he 0\lab{saddlepoint2}
\eea
The integrals of Eqns. (\ref{BFKLUC2}) and (\ref{BFKLUC3}) can be solved using the method of steepest descents to yield:
\bea
&&N_{BFKL}^\dag\Lb L - l  , \vec{b}- \vec{b}',Y - Y^{\prime}\Rb\he\lab{steepestdescents1} \\
&&\sqrt{\f{2\pi}{\bas\chi\dpr\Lb\ga_{SP}\Rb\Lb Y-Y\prm\Rb}} \,\,
n^\dag_{BFKL}(\ga_{SP}, \vec{b}-\vec{b}')\,\exp\Lb\bas\chi\Lb\ga_{SP}\Rb\Lb Y-Y\prm\Rb - \Lb L-l\Rb\,( 1 - \ga_{SP})\Rb\nn\\
&&\he\tilde{n}_{BFKL}  (\ga_{SP}, \vec{b}-\vec{b}'  )\,e^{S^\dag}\nn\\\nn\\
&&N_{BFKL}\Lb  l  , \vec{b}- \vec{b}', Y^{\prime}\Rb \he\lab{steepestdescents2}\\
&&\sqrt{\f{2\pi}{\bas\chi\dpr\Lb\ga\prm_{SP}\Rb Y\prm}} \,\,
n_{BFKL}(\ga\prm_{SP}, \vec{b}-\vec{b}')\exp\Lb\bas\chi\Lb\ga\prm_{SP}\Rb Y\prm p-  l\,(1 - \ga\prm_{SP})\Rb\nn\\
&&\he \tilde{n}_{BFKL}\Lb \ga'_{SP}, b'\Rb\,e^{S}\nn\\\nn\\
&&S\he\bas \Lb \chi(\ga')\,\,\,-\,\,\,\Lb 1 \,-\,\ga'_{SP}\Rb\,\frac{d\chi(\ga'_{SP})}{d \ga'_{SP}}
 \Rb Y^{\prime};\,\,\,\,\,\,S^\dag \he \bas\Lb \chi\Lb \gamma_{SP}\Rb -  \Lb 1 - \ga_{SP}\Rb\,\frac{d\chi(\ga_{SP})}{d \ga_{SP}}\Rb\,\Lb Y - Y^{\prime}\Rb \nn
 \eea
 where all slowly changing terms, have been absorbed by the  functions $\tilde{n}^\dag$ and $\tilde{n}$.
 \eq{SCEQ1} together with \eq{EQOM1} have the following form in the linear approximation:
 \bea
\hspace{-1.5cm}&&(1.)\,\frac{d l}{d\,Y'}  = -\,\bas\,\frac{d \chi(\ga)}{d \ga};
\,\,\,(2.)\,\om = \bas \chi(\ga);
\,\,\,\,(3.)\,
\frac{d\,\tilde{S}}{d\,Y'} =  \bas\,(1 - \ga) \,\frac{d \chi(\ga)}{d \ga} + \om;\,\,(4.)\,\frac{d\,\gamma}{d\,Y'}\,\,=\,\,0;
\label{SCEQ11}
\end{eqnarray}

It is easy to see that \eq{SCEQ11} leads to the same $S$  as \eq{steepestdescents1} and \eq{steepestdescents2}. One can see that  for $\gamma = \ga_{cr}$ for which
\beq \label{GACR}
 \chi(\ga'_{cr})\,\,\,+\,\,\,( 1 \,-\,\ga'_{cr}) \frac{d\chi(\ga'_{cr})}{d \ga'_{cr}}\,\,=\,\,0
\eeq

Then $S=0$. The equation for this line takes the form:
\beq \label{BFKUC4}
 l \,\,=\,\,\bas \frac{\chi\Lb \ga_{cr}\Rb}{ 1 - \ga_{cr}}\,Y'
 \eeq
 Repeating all the steps of the calculations above for $S^\dag$, we find that $S^\dag = 0$ on the line
 \beq \label{BFKLUC5}
 L - l\,\,=\,\,\bas \frac{\chi\Lb \ga_{cr}\Rb}{ 1 - \ga_{cr}}\,\Lb Y - Y'\Rb
 \eeq
 using the same $\ga_{cr}$ from \eq{GACR}.
  The general pattern of trajectories for the linear equation is shown in \fig{traj}.
It should be stressed that for \eq{SCEQ1} and \eq{SCEQ2}, we have different trajectories
but they are parallel and shifted by $\Delta l =\bas \Lb \chi\Lb\ga_{cr}\Rb/\ga_{cr} \Rb Y$.

\FIGURE[ht]{
\centerline{\epsfig{file=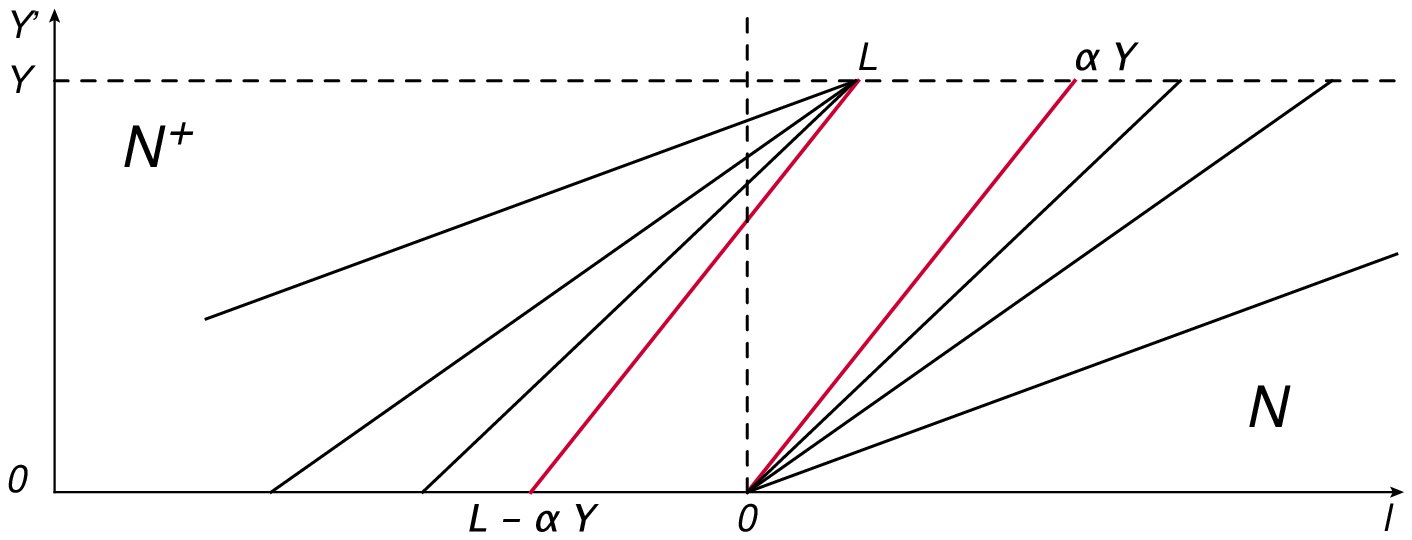,width=110mm}}
\caption{ The trajectories of the linear equations for
 $N^\dag$ and $N$: $L =\ln \Lb k^2_f/k^2_i\Rb, l = \ln\Lb k^2/k^2_i\Rb$ and $\alpha= \bas \chi\Lb\ga_{cr}\Rb/\ga_{cr}$.
 Red lines denote the critical trajectories.  Only the trajectories to the left (for $N^\dag$ ) and to the right (for $N$) are shown.}
\label{traj}}

 We need to know the initial conditions for $S$($S^\dag$) and $\ga$ ($\ga^\dag$) at $Y' =0$ ($Y'=Y$). We choose the McLerran-Venugopalan
formula\cite{MV} which is written for the dipole-target amplitude and is given in terms of $N$ as:
 \beq \label{McV}
 N\Lb r,b, Y'=0\Rb\,\,=\,\,1\,\,\,-\,\,\exp\Lb - r^2 k^2_i(b)/4\Rb
 \eeq
 where  the initial characteristic momentum $k^2_i (b) \propto T_A\Lb b\Rb$, and $T_A(b)$ is used to denote
 the number of nucleons inside the nucleus with fixed impact parameter $b$. In momentum representation $N$ is equal to (see \fig{ic}):
 \bea \label{IC}
&& N_0\Lb k, b, Y'=0\Rb\,\,\,=\,\, \int r d r \,J_0\Lb k r\Rb N\Lb r,b, Y'=0\Rb/r^2 \,\,= \,\,\h \Gamma\Big(0,\tau = \frac{k^2}{k^2_i}\Big)\,\,
\mbox{or} \,\,S\,=\,\ln\Big(\h \Gamma_0\Big(\tau\Big)\Big)\nn\\
&& \gamma_0\,\,-\,\,1\,\,=\,\,\,\frac{\partial \ln \Big( N\Lb k, b, Y'=0\Rb\Big)}{ \partial \ln \Lb k^2/k^2_i \Rb}\,\,=\,\,- e^{-\tau}\,\Big{/}\Gamma_0\Lb \tau\Rb
\eea

$ \Gamma_0\Lb \tau\Rb$ that appears in \eq{IC} is the Euler incomplete gamma function ( see formulae {\bf 8.50} in Ref.\cite{RY}).

Caution should be taken here, since
 we cannot trust the McLerran-Venugopalan formula at small dipole sizes.
Indeed, we know that in the limit of perturbative QCD, $\ga_0$ is $\ga \to 0$ contradicts this formula.
The problem is that  the simplified version of \eq{McV} does not reproduce the perturbative QCD limit at $r \to 0$.
The relation in \eq{McV} leads to $ N \propto r^2 $ at $r \to 0$,
 while the correct behaviour should be $N \propto r^2 \ln r^2$. Nevertheless we will use \eq{McV},
 because our main interest lyes  in the region in the vicinity of the saturation scale, where \eq{McV} reproduces the amplitude for $N$ quite well.
\FIGURE[ht]{
\epsfig{file=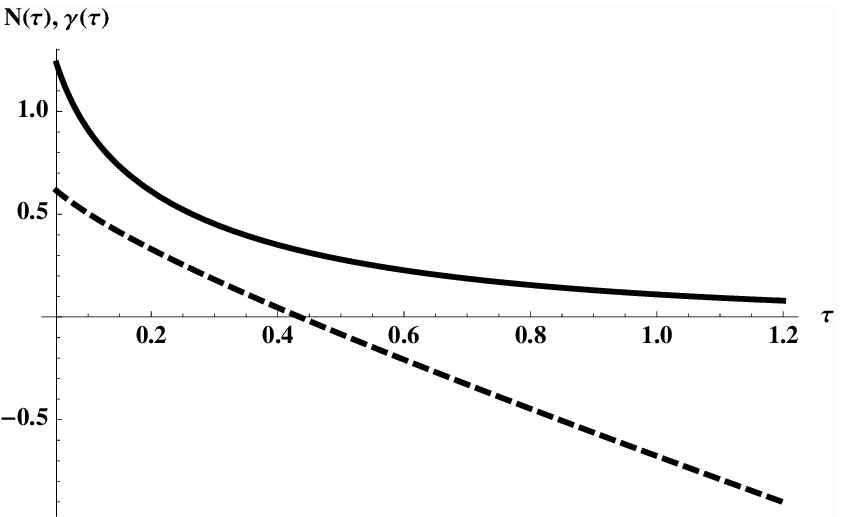,width=70mm}
\caption{Initial conditions:  $N_0(\tau)$(solid line) and $\gamma_0 $ (dotted line).}
\label{ic}}
For $N^\dag$ the initial conditions look the same, but $\tau^\dag\,=\,k^2/k^2_f$. In other words, $ N^\dag\Lb k, b, Y' =Y\Rb\,\,=\,\,e^{S^\dag_0}\,\,=\,\,\h \Gamma_0\Lb \tau^\dag\Rb$. One can see that using \eq{SCEQ1}(4) and \eq{SCEQ2}(4), we obtain that
$S^\dag\Lb l,Y'\Rb\,\,\,=\,\,\,S\Lb L,Y\Rb\,\,-\,\,S\Lb l , Y^{\prime}\Rb\,\,+\,\,S^\dag_0$.  It means that $\om^\dag\,=\,-\om$ and $ 1\,-\,\ga^+\,=\,- \,(1 - \ga)$.
It is easy to see that the system of equations in Eqns. (\ref{SCEQ2}) degenerate to the system of equations in (\ref{SCEQ1}).

\subsection{The final system of equations}
Therefore, instead of \eq{SCEQ1} and \eq{SCEQ2} , we can solve the following system of equations.
\bea \label{SCEQF}
&&(1.)\,\,\,\,\,\,\,\frac{d l}{d\,t}\,\,=\,\,-\,\bas\,\frac{d \chi(\ga)}{d \ga},
\,\,\,\,\,\,\,\,\,\,\,\,(2.)\,\,\,\,\,\,\,
\frac{d\,Y^\prime}{d\,t}\,\,=\,\,\,1,\,\,\,\,\,\,\,\nn\\
 &&(3.)\,\,\,\,\,\,\,
\frac{d\,S}{d\,t}\,\,=\,\,  \bas\,\Lb 1 \,-\,\ga\Rb\,\frac{d \chi(\ga)}{d \ga}\,\,+\,\,\om,\nonumber \\
&&(4.)\,\,\,\,\,\,\,\frac{d\,\gamma}{d\,t}\,\,=
\,\,\,-\,\bas \kappa\Lb \ga\Rb\,(1 - \ga)\,e^{S\Lb
L,Y\Rb \,\,-\,\,S\Lb l, Y'\Rb \,+\,S_0 }\,\,+\,\,\bas\,(1\,-\,\ga)\,e^{S\Lb l, Y'\Rb},\nn\\
&& (5.)\,\,\,\,\,\,\,
\frac{d\,\omega}{d\,t}\,\,=\,\,\,\,\bas\Big( \om
\kappa\Lb \ga\Rb\,e^{S\Lb
L,Y\Rb \,\,-\,\,S\Lb l, Y'\Rb \,+\,S_0 }\,\,-\,\,\om\,e^{S\Lb l, Y'\Rb}\Big),
\eea

For the sake of simplicity, the label \,$\widetilde{}$\, has been omitted everywhere in \eq{SCEQF}.
\eq{SCEQF} can be re-written in a different form,  namely
\bea \label{SCEQFF}
&&(1.)\,\,\,\,\,\,\,\frac{d l}{d\,Y'}\,\,=\,\,-\,\bas\,\frac{d \chi(\ga)}{d \ga},
\,\,\,\,\,\,\,\,\,\,\,\,(2.)\,\,\,\,\
\frac{d\,S}{d\,Y'}\,\,=\,\,\,  \bas\,\Lb 1 \,-\,\ga\Rb\,\frac{d \chi(\ga)}{d \ga}\,\,+\,\,\om,\,
\,\,\,\,(4.)\,\,\,\,\,\,\,\frac{d\,\gamma}{d\,\om}\,\,=\,\,-\,\frac{1 - \ga}{\om},\nn\\
&& (5.)\,\,\,\,\,\,\,\frac{ d \ga}{d S}\,\,=\,\,\frac{ -\kappa\Lb \ga\Rb\,\,e^{S\Lb
L,Y\Rb \,\,-\,\,S\Lb l, Y'\Rb \,+\,S_0 }\,\,+\,\,e^{S\Lb l, Y\,\Rb}}{d \chi\Lb \ga
\Rb/d\ga\,+\,\,\om/(1 - \ga)}
\eea

\FIGURE[h]{\begin{tabular}{c c c}
\epsfig{file=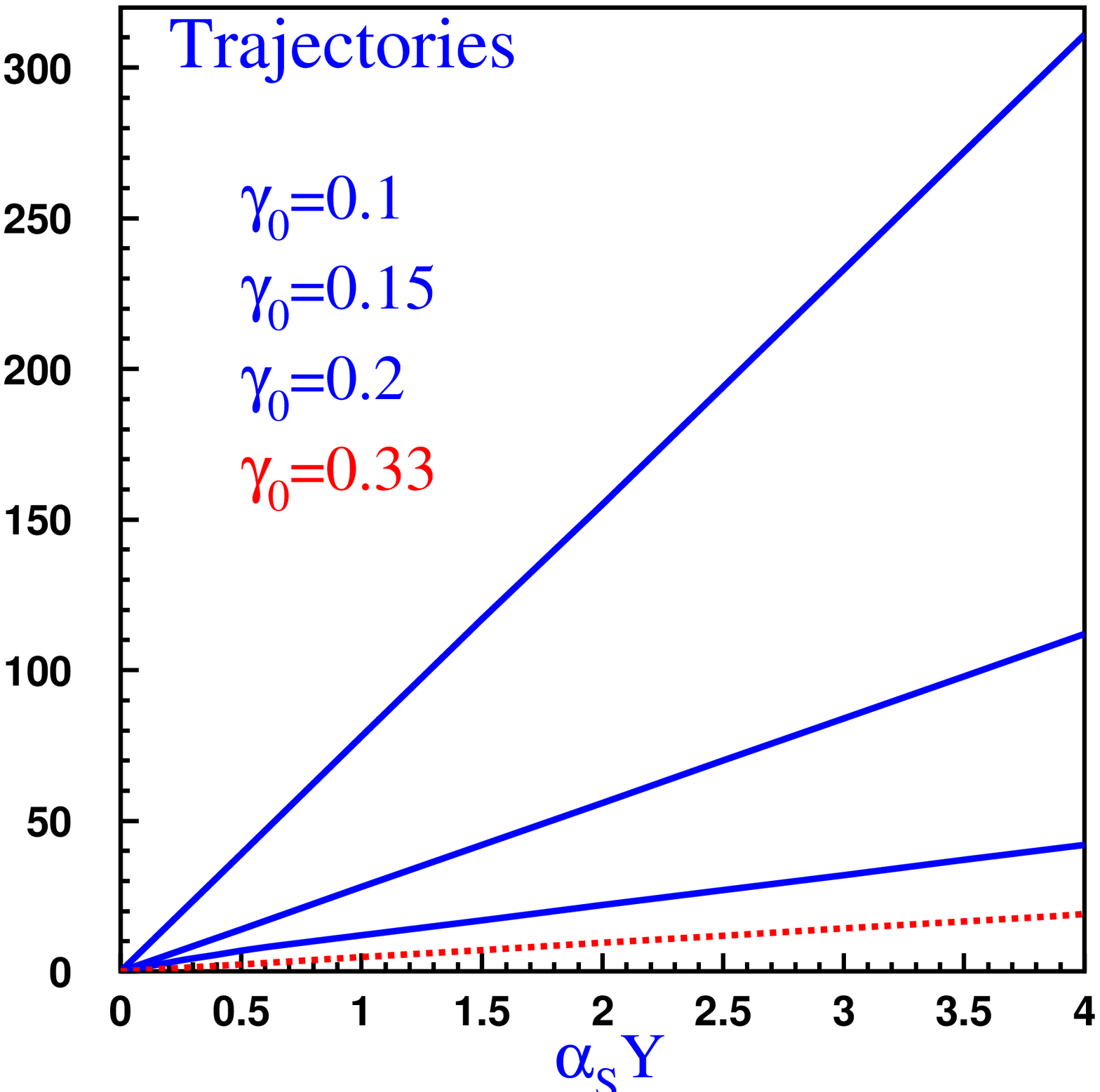,width=50mm}& \epsfig{file=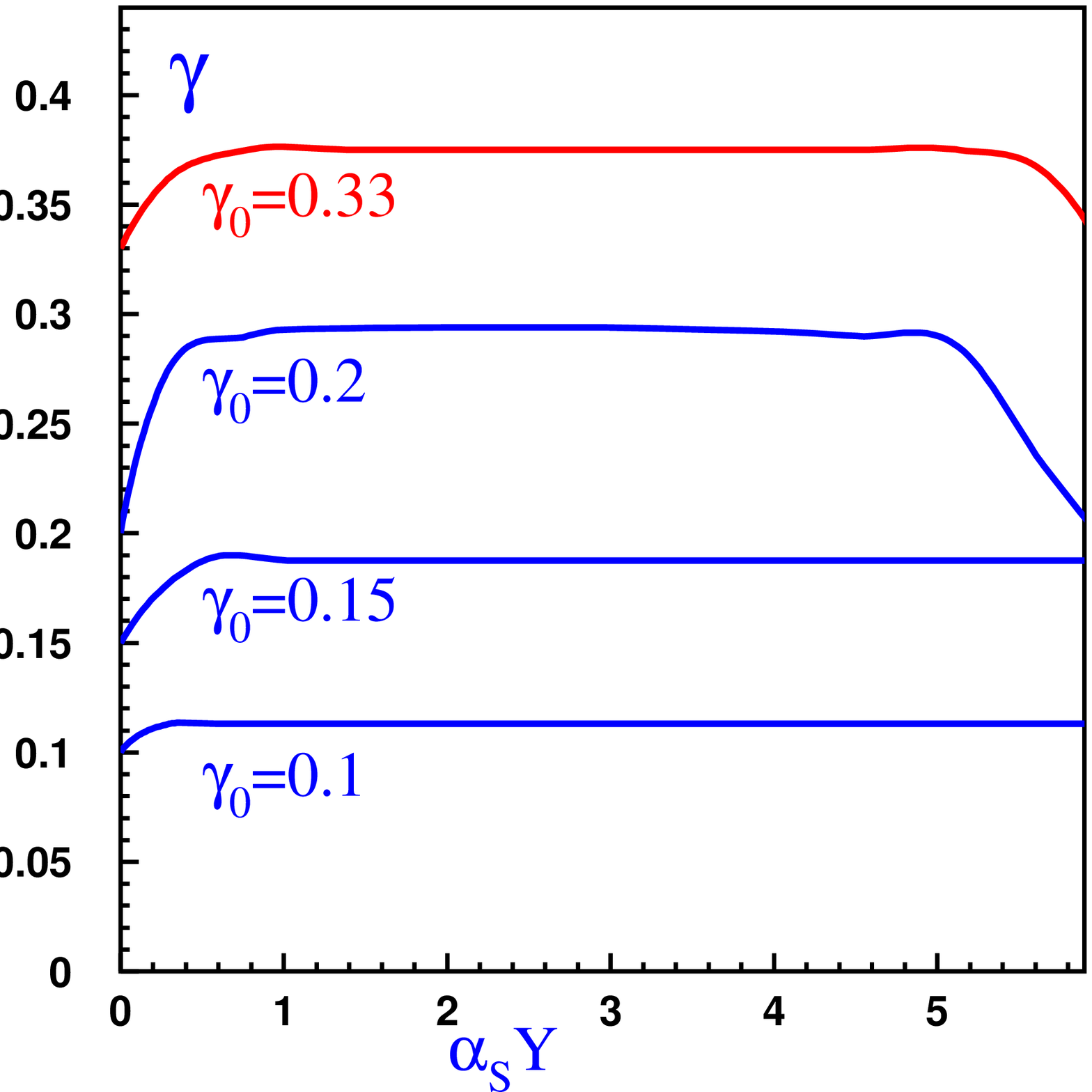,width=50mm}&\epsfig{file=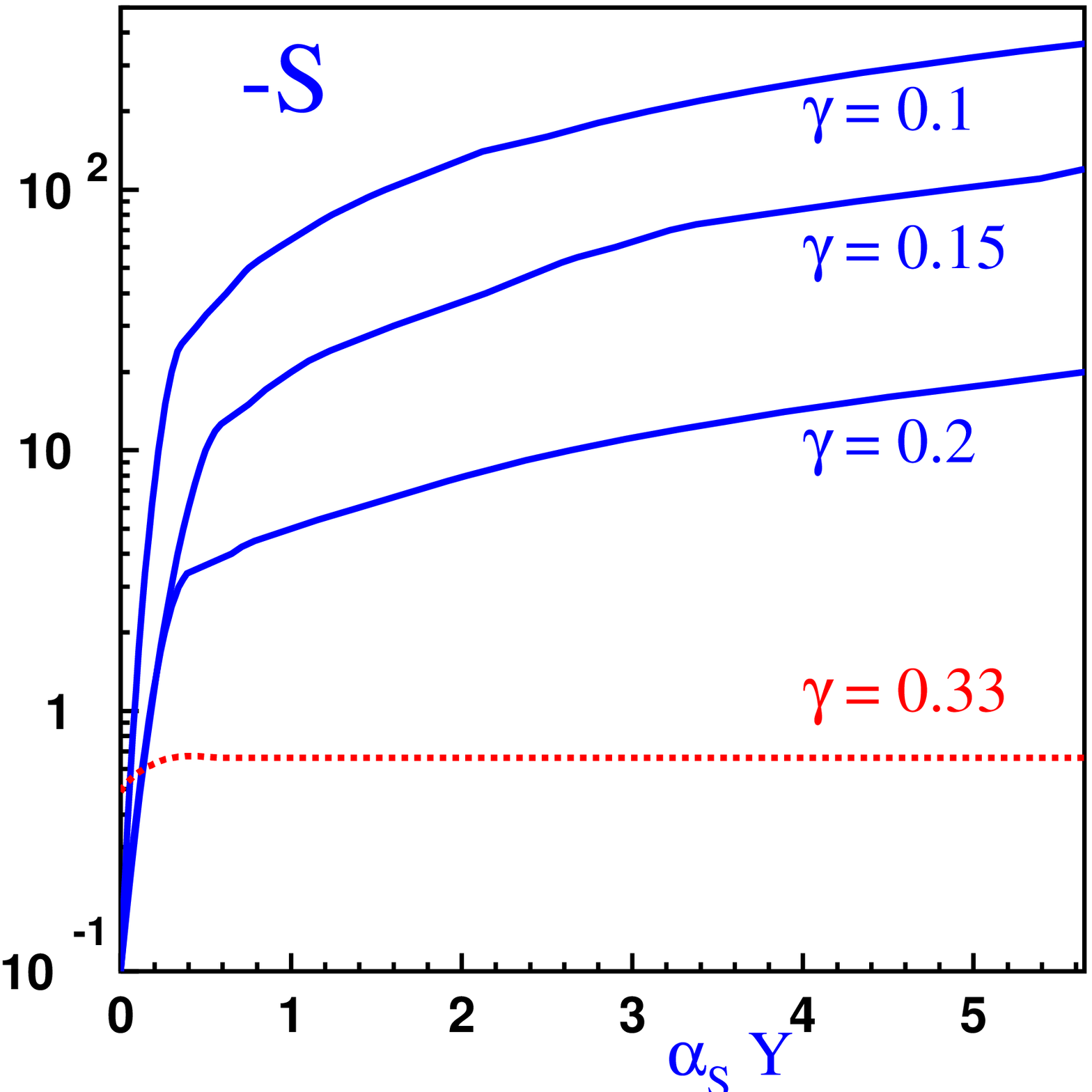,width=50mm}\\
\fig{soltr}-a & \fig{soltr}-b& \fig{soltr}-c\\
\end{tabular}
 \caption{Semiclassical solution: trajectories $l\Lb Y\Rb$ (\fig{soltr}-a), $\ga$ versus $\bas Y$ (\fig{soltr}-b ) and  $S$ versus $\bas Y'$(\fig{soltr}-c). The solution with $\ga$ close to $\ga_{cr}$ is shown in red by dotted line.
$\bas Y$ is chosen to be equal 6.
 }
 \label{soltr}}
\eq{SCEQFF}-4 has the solution   $ \om =\,\bas\, Const ( 1 - \ga)$ where $Const$ can be determined from the initial conditions.
\beq \label{Const}
\mbox{Const}\,\,=\,\,\,\frac{\chi\Lb \ga_0\Rb - \kappa\Lb \ga_0\Rb\, e^{S\Lb L,Y\Rb} \,-\,e^{S_0}}{1 - \ga_0}
\eeq

Notice that the value of this constant depends on $S\Lb L,Y\Rb$, but for large and negative  $S\Lb L,Y\Rb$
which we are dealing with, for $\ga_0 < \ga_{cr}$, we can safely neglect this dependence.
Introducing
 $\hat{S}\,\,= S\Lb l,Y'\Rb - \Lb S\Lb L,Y\Rb + S_0\Rb\Big{/}2$ we can rewrite \eq{SCEQFF}-5 in the form

\beq
\label{EQ5}
\frac{ d \ga}{d S}\he e^{
\Lb S\Lb L, Y\Rb + S_0\Rb/2}\,\,\,\,\Lb\frac{ -\kappa\Lb \ga\Rb\,\,e^{-\,\,\hat{S}\Lb l, Y'\Rb  }\,\,+\,\,e^{\hat{S}\Lb l, Y'\,\Rb}}{d \chi\Lb \ga
\Rb/d\ga\,+\,\,\,Const}\Rb
\eeq

\FIGURE[h]{
\epsfig{file=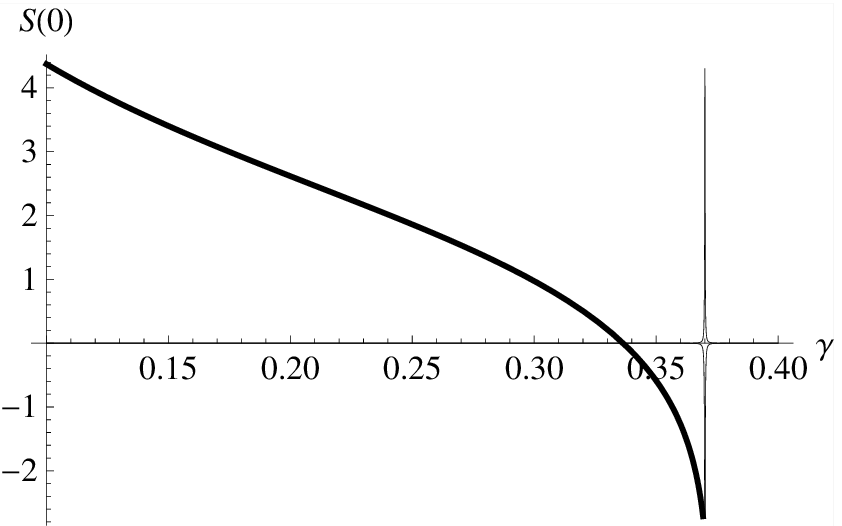,width=60mm}
\caption{Dependence of $S\Lb Y'=0\Rb = S_0$ on $\ga_{cr}$ from \eq{GACRF} (thick line) and $ \ga_{cr} = 0.37$ (solution of \eq{GACR}, thin line).}
\label{gacr}}

We follow the following strategy for solving this equation. First we assume that we are looking for the solution in the interval
 $\{\hat{S}_0, \hat{S}_{max}\}$,
where we have selected $\hat{S}_{max}$.
 Using the initial conditions, we can calculate $\hat{S}\Lb l, Y=0\Rb$ through $S_0 $ and $S_{max}$, and specify the coefficient in front of \eq{EQ5}.
  Solving \eq{SCEQFF}-2 we will find  $Y'$ as a function of $\hat{S}$. In particular $Y'\Lb S_{max}\Rb \neq Y$,
 however trying several times we find a value for $ S_{max}$ that will give $Y'\Lb S_{max}\Rb  $. \fig{soltr} shows the solutions for four  values of $\ga_0$ that are smaller than $\ga_{cr}$.
One can see that the trajectories and the values of $S$ and $\ga$ are close to the solution of the linear equation. Indeed, $dS/dY' \,<\,0$ for  $\ga < \ga_{cr}$ and
 therefore, the value of $S$ decreases due to evolution. Starting with $S_0 <0$ one can see that the contribution of the non-linear terms becomes less and less at higher
 values of $Y'$ than at $Y'=0$. Therefore, only at small values of $Y'$ we can see a deviation from the solution to the linear equation,
 as shown in \fig{soltr} (see \fig{soltr}-b for example).
\eq{SCEQFF}-4 has the form \beq
\label{EQ51}
\frac{ d \ga}{d Y'}\,\,=\,\,\bas\Lb 1-\ga\Rb\,e^{
\Lb S\Lb L, Y\Rb + S_0\Rb/2}\Big\{-\kappa\Lb \ga\Rb\,\,e^{-\,\,\hat{S}\Lb l, Y'\Rb  }\,\,+\,\,e^{\hat{S}\Lb l, Y'\,\Rb}\Big\}
\eeq and due to the smallness of the factor $\exp\Lb\Lb S\Lb L, Y\Rb + S_0\Rb/2\Rb$, then $ d \ga/d Y' $ turns out to be small at large $Y'$, leading to a constant $\ga$, as shown in \fig{soltr}-b.
The value of $\ga_{cr}$ is slightly different from the one from \eq{GACR}, due to a contribution from
 the non-linear terms to \eq{SCEQFF}-3, and it depends on the initial condition for $S\Lb Y'=0\Rb = S_0$.
 Indeed, for the trajectory on which $S\Lb l,Y'\Rb \,=\,S_0$ is constant we have the following equation: \beq \label{GACRF}
(1 - \ga_{cr})\frac{d \chi\Lb \ga_{cr}\Rb}{d \ga_{cr}} \,+\,\chi\Lb \ga_{cr}\Rb\,\,-\,\,\Lb \kappa\Lb \ga_{cr}\Rb \,+\,1\Rb e^{S_0}\,\,=\,\,0
\eeq The dependence of $S_0$ on $\ga_{cr}$ from \eq{GACRF} is shown in \fig{gacr}. One can see that for $S_0 < 0$ ($N_0  < 1$),
 the value of $\ga_{cr}$ is close to the solution to \eq{GACR}.
From \fig{soltr} one can see that we start the evolution from the value of $\ga$ that is close to $\ga_{cr}$, 
$\ga$ steeply increases to $\ga=\ga_{cr}$ and freezes at this value leading to constant $S$ almost in the entire kinematic region of $\bas Y'$.
 For $\ga  >  \gamma_{cr}$, the solution will lead to $S\Lb l,Y'\Rb $
 that increases with $Y'$, and we need to search for a different method  of finding the solution, other than the semi-classical approach.

\section{Solution inside the saturation domain}
\subsection{General consideration}

As discussed above, the semi-classical approach cannot be used inside of the saturation region.
 It should be mentioned that at large $l$, the amplitude behaves as $ \h l$, which is certainly not the function for which we can use the semi-classical approach.
 However, it has been noticed in Ref.\cite{BKL} that introducing new functions : $\phi\Lb l',Y', b\Rb$ and $\phi^\dag\Lb l',Y', b\Rb$ instead of $N$ and $N^\dag$, defined as
\beq \label{NEWF}
N\Lb l,Y', b\Rb\,\,=\,\,\h \int_l\,d l' \Big( 1 \,-\,e^{- \phi\Lb l',Y', b\Rb}\Big),\,\,\,\,\,\,\,\,\,
N^\dag\Lb l,Y', b\Rb\,\,=\,\,\h \int_l\,d l' \Big( 1 \,-\,e^{- \phi^\dag\Lb l',Y', b\Rb}\Big)
\eeq

Then we can indeed  use the semi-classical approach for these functions. The first observation is that,
 from the property of \eq{HSC} and the definition of \eq{BFKLUC3}:
 
\bea \label{SH}&&
H\,\frac{\partial N\Lb l ,Y', b\Rb}{\partial l}\he\int \frac{d \ga}{2 \pi i}\,\bas\chi\Lb \ga \Rb\,\ga \,n\Lb \Lb \ga\,-\,1\Rb\,, Y', b\Rb\,
\exp\Lb\bas\chi\Lb\ga\Rb Y' -  (1 - \ga)\, l\Rb\,\nn\\
&&\equiv\hspace{0.3cm}\bas\,\int  \frac{d \ga}{2 \pi i}\Lb\chi\Lb \ga\Rb - \frac{1}{1 - \ga}\Rb\,\Lb\ga\,-\,1\Rb\, n\Lb \ga, Y', b\Rb\,
\exp\Lb \bas\chi\Lb\ga\Rb Y' - (1 -  \ga)\, l\Rb\,\,-\,\,\bas N\Lb l, Y', b\Rb\nn\\
&&\equiv\hspace{0.3cm}\bas\,{\cal L}\Lb - \frac{\partial}{\partial l}\Rb\,\frac{\partial N\Lb l ,Y', b\Rb}{\partial l}
\,\,-\,\,\bas \,N\Lb l, Y', b\Rb\nn
\eea

The second observation is related to the function ${\cal L\Lb \ga \Rb}$, namely that  its expansion with respect to $(1 - \ga)$ starts with $( 1 - \ga)^2$ as the first non-zero term. That is:

\beq \label{LEXP}
{\cal L}\Lb - \frac{\partial}{\partial l}\Rb\,\,=\,\,- \frac{d^2 \psi(z)}{d z^2}|_{z = 1} \frac{\partial^2}{\partial l^2}
\,\,\,- \,\,\frac{1}{12} \frac{d^4 \psi(z)}{d z^4}|_{z = 1}\frac{\partial^4}{\partial l^4}\,\,\,-\,\,\dots
\eeq
where $\psi(z) = d  \ln \Ga(z) / d z$ is the Euler $\psi$-function (see Ref.\cite{RY}).
 Following the definition of \eq{NEWF} and
assuming that $\phi$ and $\phi^\dag$ are smooth functions, we can replace
\footnote{For $\partial \phi/\partial l$ ($\partial \phi^\dag/\partial l$)   we use the notations $\ga_{\phi}$ and $\ga^\dag_\phi$. We hope that it will not
 lead to any confusions with the notation,  due to the similarity with $\ga$ and $\ga^\dag$, that was heavily used in the previous section.}
\bea
&&\Lb\f{\D}{\D l}\Rb^nN\Lb l,Y', b\Rb = -\h \Lb\f{\D}{\D l}\Rb^{n-1}\Lb 1 \,-\,e^{- \phi\Lb l,Y', b\Rb}\Rb  = \h\Lb -\f{\D\phi}{\D l}\Rb^{n - 1}\,e^{- \phi\Lb l,Y', b\Rb}
\hspace{1cm}\lab{NEWFderivative1}\\
&&\Lb\f{\D}{\D l}\Rb^nN^\dag\Lb l,Y', b\Rb =
-\h \Lb\f{\D}{\D l}\Rb^{n-1}\Lb 1 \,-\,e^{- \phi^\dag\Lb l,Y', b\Rb}\Rb = \h\Lb -\f{\D\phi^\dag}{\D l}\Rb^{n - 1}\,e^{- \phi^\dag\Lb l,Y', b\Rb}
\hspace{1cm}\lab{NEWFderivative2}\eea

Inserting  Eqs. (\ref{LEXP}), (\ref{NEWFderivative1}) and (\ref{NEWFderivative2}) and  into \eq{EQOM1} and \eq{EQOM2}, and introducing the notation
$\partial \phi/\partial Y'\,=\,\om_\phi$ and $\partial \phi^\dag/\partial Y'\,=\,\om^\dag_\phi$ 
we obtain:

\bea
&&0\he \Big( \om_\phi\,\,+\,\bas \,{\cal L}\Lb - \ga_\phi\Rb\Big) \Lb 1 - \ga_\phi\Rb^2 \ga_\phi e^{- \phi \Lb l,b,Y'\,\Rb}\,\,+\,\,2\,\bas  \Lb\frac{\partial}{\partial l}\,+\, 1\Rb^{2}\,\frac{\partial}{\partial l}
\Lb\,e^{ - \phi \Lb l,b,Y'\,\Rb} \,N\Lb l,b,Y'\,\Rb\,\Rb\hspace{1cm}\lab{SDEQ1}\\
&  & +\,\,\bas \,2\,N^\dag\Lb l,b,Y\prm\Rb \Lb\frac{\partial}{\partial l}\,+\,1\Rb^{2}\frac{\partial}{\partial l} \,e^{- \phi\Lb l,b,Y\prm\Rb} \nn\\
& & \nn\\
&&0\,\,=  \, \Big(- \om^\dag_\phi\,\,+\,\bas \,{\cal L}\Lb - \ga^\dag_\phi\Rb\Big)\Lb 1 - \ga^\dag_\phi\Rb^2 \ga^\dag_\phi e^{- \phi^\dag \Lb l,b,Y'\,\Rb}\,\,+\,\,2\,\bas  \Lb\frac{\partial}{\partial l}\,+\, 1\Rb^{2}\,\frac{\partial}{\partial l}
\Big\{e^{ - \phi^\dag \Lb l,b,Y'\,\Rb} \,N^\dag\Lb l,b,Y'\,\Rb\Big\}\hspace{1cm}\lab{SDEQ2}\\
&  & +\,\,\bas \,2\,N\Lb l,b,Y\prm\Rb \Lb\frac{\partial}{\partial l}\,+\,1\Rb^{2}\frac{\partial}{\partial l} \,e^{- \phi^\dag\Lb l,b,Y\prm\Rb} \nn
\eea

 One can see that

 \bea \label{SDEQ3}
&&    \Lb\frac{\partial}{\partial l}\,+\, 1\Rb^{2}\,\frac{\partial}{\partial l}
\Big\{e^{ - \phi \Lb l,b,Y'\,\Rb} \,N\Lb l,b,Y'\,\Rb\Big\} \he e^{- \phi}\Big\{ \h\Lb 1 - 6\ga_\phi + 7 \ga^2_\phi\Rb\,e^{- \phi}\,\,-\,\,\,\h\Lb 1 \ - 4 \ga_\phi\,+\,3 \,\ga^2_\phi \Rb \\
 &&-\,\,\ga_\phi (1 - \ga_\phi)^2\, N\Lb l,b,Y'\,\Rb \Big\}\nn
 \eea

 Dividing both sides of \eq{SDEQ1} by $\bas \, e^{- \phi}$ and introducing the new variable $\bar{\om} \,=\,\om/\bas$, which corresponds to the
change $Y' $ to $ \bar{Y}' \,=\,\bas Y'$ we obtain:

 \bea \label{SDEQ4}
\hspace{-0.3cm}&& - \bar{\om}_\phi  - {\cal L}\Lb - \ga_\phi\Rb  + 2\,N^\dag\Lb l, b, Y'\Rb -  \frac{\Lb 1 - 6\ga_\phi + 7 \ga^2_\phi\Rb\,e^{-  \phi}\,-\,\Lb 1 - 4 \ga_\phi +3 \,\ga^2_\phi \Rb }{ ( 1 - \ga_\phi)^2 \ga_\phi}
 + 2 N\Lb l,b,Y'\,\Rb \,=\,0
 \eea

A similar equation can be written for
$\phi^\dag$. However,  we assume that $\phi^\dag\Lb l,b,Y'\,\Rb\,\,=\,\,\phi\Lb L - l,b,Y - Y'\,\Rb $ based on our experience with the semi-classical solution.
 Looking for the solution that has a geometric scaling behaviour \cite{GS}, we expect that $\phi\Lb l, b, Y'\Rb$ is a function of one  variable:
  \beq \label{ZSOL}
  z\,\,=\,\,\ln\Lb Q^2_s\Lb Y',b\Rb/k^2\Rb\,\,\,=\,\,\,\frac{\chi\Lb \ga_{cr}\Rb}{1 - \ga_{cr}}\,Y'\,\,-\,\,l
  \eeq

  \subsection{Asymptotic solution}
  Plugging in  $\phi^\dag\Lb l,b,Y'\,\Rb\,\,=\,\,\phi\Lb L - l,b,Y - Y'\,\Rb $ and the geometric scaling behavior of \eq{SDEQ4} that we anticipate,
  and assuming that $\phi \,\gg\,1$ we obtain:
  \bea\label{SDEQ5}
    &&G\Lb \tilde{\ga}\Rb\,=\,  z_Y\\
    &
    \mbox{where} &\,\,\,\,G\Lb \tilde{\ga}\Rb \,=\, \frac{\chi\Lb \ga_{cr} \Rb}{1 - \ga_{cr}}\, \tilde{\ga}
    \,+\, {\cal L}\Lb \tilde{\ga}\Rb
    +\frac{1 + 4  \tilde{\ga} +3  \tilde{\ga}^2 }{ ( 1 + \tilde{\ga})^2 \,\tilde{\ga}}\nn\\
    &~&  z_Y\,\,\equiv\,\ln\Lb Q^2_s\Lb Y\Rb/k^2_i\Rb\,\,\,=\,\,\,\frac{\chi\Lb \ga_{cr}\Rb}{1 - \ga_{cr}}\,Y\,\,-\,\,L  \,\nn
     \eea

It should be noticed that  $\tilde{\ga}$ is defined as $\tilde{\ga}\,=\,d \phi/d z$.
The function $G\Lb \ga_\phi\Rb$ is shown in \fig{solsd}. One can see that we have  three types of solutions to \eq{SDEQ5}:

\FIGURE[h]{\begin{tabular}{c  c}
\epsfig{file=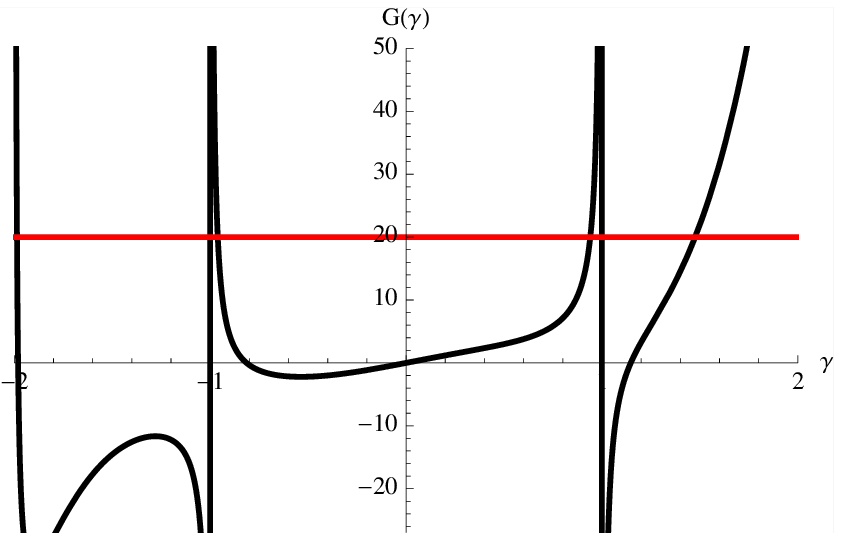,width=70mm}& \epsfig{file=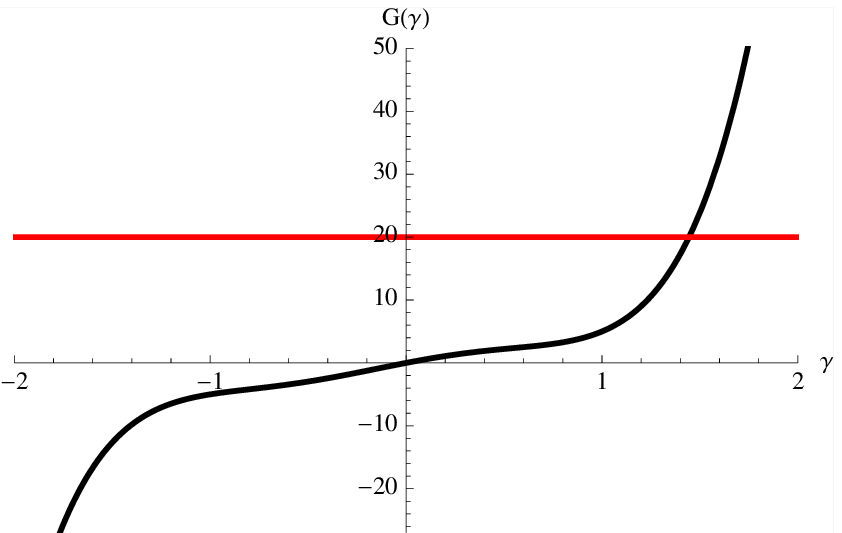,width=70mm}\\
\fig{solsd}-a & \fig{solsd}-b\\
\end{tabular}
 \caption{Function $G\Lb \ga \Rb$ (solid line)  versus $\ga$. The red line shows the r.h.s. of  \protect\eq{SDEQ5} at $z_Y = 20$. \fig{solsd}-a shows  $G(\ga)$ as it is written in \protect\eq{SDEQ5}, while  $G(\ga)$ with ${\cal L}=0$ is plotted in \fig{solsd}-b.}\label{solsd}}

\begin{enumerate}
 \item \quad At large $\tilde{\ga}$ then $G\Lb \tilde{\ga}\Rb  \to \chi(\ga_cr)/( 1 - \ga_{cr})\,\tilde{\ga}$ and we have the solution:
$\tilde{\ga}\,\,=\,\, (1 - \ga_{cr})/\chi\Lb \ga_{cr}\Rb\,z_Y$ which translates into $\phi \,=\,\Lb (1 - \ga_{cr})/\chi\Lb \ga_{cr}\Rb\Rb\,z_Y\,z$;

    \item \quad  We have solutions at $\tilde{\ga}\,\to n$ where $n\,=\,1,2,\dots $ which lead to $\phi = n z$.
    We only need to take into account $n=1$, since other values of $n$ give smaller contributions at large $z$. In vicinity $\ga \to 1$ $G\Lb \ga \Rb \,= - 1/(1 - \ga)$ and the solution to \eq{SDEQ5} gives $\phi = \Lb 1  + 1/z_Y\Rb z$;
 \item \quad  Solutions where $ \tilde{\ga}\,\to - n $ we do not consider, since they lead to decreasing $\phi$ at large $Y$.
 \end{enumerate}

 It is interesting to notice, that at in the case where we restrict ourselves  to  the leading twist contributions to the BFKL kernel \cite{LT},
 only the first solution survives (see \fig{solsd}-b).

Using \eq{NEWF} we can obtain the asymptotic solution for the scattering amplitude, namely
\bea 
&& N\Lb z_Y\Rb\,\,=\,\,\h \int^{z_Y}_0\,d z \Big( 1 - \exp\Lb - \Lb1 - \ga_{cr})/\chi\Lb \ga_{cr}\Rb\Rb\,z_Y\,z\Rb\Big)\,\,\,\longleftarrow\,\,\mbox{leading twist};\label{AMF1}\\
&& N\Lb z_Y\Rb\,\,=\,\,\h \int^{z_Y}_0\,d z \Big( 1 - \exp\Lb -\,z\Rb\Big)\,\,\,\longleftarrow\,\,\mbox{general asymptotic behaviour};\label{AMF2}
\eea

It should be noticed that the expression in parentheses $\Big( \Big)$ acxtually gives the scattering amp-litude in the coordinate representation.
\subsection{General equations}
        In general, \eq{SDEQ4} can be reduced to the differential equation by taking derivatives with respect to $z$ on both sides of the equation.
  The resulting expression is:

        \beq \label{SDEQ6}
        \frac{d G\Lb \tilde{\ga}\Rb}{ d \tilde{\ga}}\,\frac{d \tilde{\ga}}{d z}\,\,=\,\,e^{ - \phi\Lb z\Rb}\,\,-\,\,e^{ - \phi\Lb z_Y \,-\, z\Rb}     ,   \,\,\,\,\,\,\,\,\,\,\frac{d \phi\Lb z \Rb}{ d z}\,\,=\,\,\tilde{\ga},
        \eeq

        This equation belongs to the class of delay differential equations, and  the solution to this equation we hope to publish 
in an upcoming paper,
 since it is a separate and rather difficult problem beyond the scope of this paper (see for example Ref.\cite{HAL}).

        \section{Conclusions}
        In conclusion, we summarize our results as follows. First we re-wrote the action of the BFKL Pomeron calculus, and we derived 
the equations in momentum representation.
 It turns out that the  equations that we obtain have  a simpler form, than in coordinate representation in the format that they were originally derived in Ref.\cite{BRA}.
 Second, we found the semi-classical solution to these equations, to the right of the critical line . In the saturation domain,
 we reduced these equations to the class of delay differential equations, and we found their asymptotic solution.
 This solution shows, that the nucleus-nucleus amplitude in coordinate representation  approaches unity as $ 1 - \Delta N$, where $\Delta N \propto \exp\Lb - z_Y\Rb$,
 where  $z_Y\,\,\equiv\,\ln\Lb Q^2_s\Lb Y\Rb\,r^2_i\Rb     $ , where $r_i$ is the size of the colourless dipole in the nucleus.
If we take into account only the leading twist part of the BFKL kernel, the behaviour of $\Delta N $ is similar to the behaviour of the amplitude
 of the dilute-dense parton system interaction, given in Ref. \cite{LT}.

        We hope that this paper will show, that the problem of the nucleus-nucleus interaction in the framework of the BFKL Pomeron calculus,
 can be solved. We hope that this development will motivate further efforts towards understanding the dense-dense scattering system.

\section{Acknowledgements}

This research was supported by the Funda\c{c}$\tilde{a}$o para ci$\acute{e}$ncia e a tecnologia (FCT),
and CENTRA - Instituto Superior T$\acute{e}$cnico (IST), Lisbon and  by the  Fondecyt (Chile) grants 1100648, 1095196 and  DGIP 11.11.05. 
One of us (JM) would like to thank Tel Aviv University for their hospitality on this visit,
during the time of the writing of this paper.

\end{document}